\documentstyle[preprint,aps]{revtex}
\tightenlines
\begin{document}
\draft

\title{Pattern Formation in Quantum Turing Machines} 

\author{Ilki~Kim\footnote{E-mail: ikim@theo.physik.uni-stuttgart.de} 
        and G\"{u}nter~Mahler}

\address{Institut f\"{u}r Theoretische Physik, Universit\"{a}t Stuttgart\\ 
         Pfaffenwaldring 57, 70550 Stuttgart, Germany} 

\date{\today}

\maketitle

\begin{abstract}
 We investigate the iteration of a sequence of local and pair unitary 
 transformations, which can be interpreted to result from a Turing-head 
 (pseudo-spin $S$) rotating along a closed Turing-tape ($M$ additional 
 pseudo-spins). The dynamical evolution of the Bloch-vector of $S$, which 
 can be decomposed into $2^{M}$ primitive pure state Turing-head trajectories, 
 gives rise to fascinating geometrical patterns reflecting the entanglement 
 between head and tape. These machines thus provide intuitive examples for 
 quantum parallelism and, at the same time, means for local testing of 
 quantum network dynamics. 
\end{abstract}


\narrowtext


Despite the lack of a clear-cut definition, 
the physics of {\it complexity} \cite{LIC83} has intrigued physicists for 
many years. For continuous classical systems with 
few degrees of freedom the notion of chaos has attracted much interest as a 
sign of uncontrollability \cite{HIL94}. For discrete classical systems in the 
form of 
cellular automata the notion of computational irreducibility has been 
introduced to account for the lack of ``short-cuts'', i.e. our inability to 
predict the respective state evolution without following the detailed 
dynamics step by step \cite{WOL85}. The linearity of quantum dynamics 
appears to make the 
respective evolution ``well-behaved'' from the start. The limit of control, 
nevertheless, abounds even for modestly large quantum networks \cite{MAH95} 
due to the, typically, exponentially large Hilbert-space, in which the state 
evolves \cite{FEY82}. It has 
been shown that if this kind of ``quantum-complexity'' could be harnessed, 
new efficient modes of computation should become available \cite{DIV95,EKE96}. 
However, one will 
first have to find ways to circumvent that disastrous exponential blow-up. 

A quantum network (composed of $N$ subsystems) is defined by its 
Hamiltonian-operator $\hat{H}$. This, $\hat{H}$, is also the generator of the 
respective unitary (system-) dynamics, $\hat{U}(t)=\exp{(-i\hat{H}t/\hbar)}$, 
which transforms a given initial state $|\psi_{0}>$ into a final state 
$|\psi'>$ after some given time $\Delta t$: We thus have a one-parameter 
transformation $\hat{U}(\Delta t)$ operating on arbitrary initial states 
(requiring a number of state parameters which grows exponentially with $N$). 
To improve control it is therefore tempting to consider, instead, arbitrary 
unitary transformations acting on one given initial state: 
In fact, this type of 
scenario underlies most current quantum-computational schemes \cite{EKE96}. 

Any system-dynamics can be approximated as an iterative sequence of 
unitary basis operators (so-called ``gates'' \cite{LLO96}). In this letter we 
address a quantum Turing machine (QTM) architecture 
\cite{DEU85,DEU89,BEN82,BEN96,BEN98} which 
can be understood as a specific and formalized version of such an iterative 
map. Typically, one will be unable to ``observe'' the network in full detail; 
one then usually resorts to ``macro-observables''. 
Here we focus, instead, on a 
single microscopic subsystem, the ``Turing-head'' $S$. To predict its state 
exactly, the full network state is required, though. However, while the 
evolution of arbitrary initial states by a given map seems exponentially 
``hard'', the evolution of some specific initial state by a whole class of 
maps turns out to be ``easy'' and is not at all limited to small 
$N$-networks. Furthermore we will show that the evolution of the Turing-head 
in its reduced space gives rise to geometrical patterns reflecting the 
entanglement between Turing-head and Turing-tape. These patterns can be 
thought to result from the superposition of exponentially many ``basic'' 
Turing-machines, an intuitive example of ``quantum parallelism''. 


The quantum network to be considered here is composed 
of $N$ ($=M+1$) pseudo-spins 
$|j(\mu)>$, $j=0,1$; $\mu=S,1,2, \cdots, M$ (Turing-head $S$, 
Turing-tape 
spins $1,2, \cdots, M$) so that its network-state $|\psi>$ lives in the 
{$2^{M+1}$-$\dim$ensional} Hilbert-space spanned by the product wave-functions 
$|j(S) k(1) \cdots l(M)>\, = |jk \cdots l>$. Correspondingly, any (unitary) 
network-operator can be expanded as a sum of product-operators. 
The latter may be based on the following traceless $SU(2)$-generators 
\begin{eqnarray} 
\label{lambda_s}
\hat{\lambda}_{x}(\mu) & = & \hat{P}_{01}(\mu) + \hat{P}_{10}(\mu) \nonumber \\
\hat{\lambda}_{y}(\mu) & = & i \hat{P}_{01}(\mu) - i \hat{P}_{10}(\mu)  \\
\hat{\lambda}_{z}(\mu) & = & \hat{P}_{11}(\mu) - \hat{P}_{00}(\mu)\,, 
\nonumber 
\end{eqnarray}
where $\hat{P}_{ij}(\mu) = |i(\mu)><j(\mu)|$ is a (local) transition operator. 

We now consider the iterative map for which 
each full cycle $p=1,2, \cdots$\, consists of a sequence of\, $2M$ unitary 
transformations $\hat{U}_{n}$, $n=1,2, \cdots, 2M$. 
At step $m$, $m=n+2M(p-1)$, we thus have
\begin{equation}
\label{psi_m}
|\psi_{m}> = \hat{U}_{n} \cdots \hat{U}_{2} \hat{U}_{1} \bigl(\hat{U}_{2M} 
\cdots \hat{U}_{2} \hat{U}_{1}\bigr)^{p-1}\,|\psi_{0}>\,.
\end{equation}
Presently we identify the $\hat{U}_{n}$ with the local unitary transformation 
on the Turing-head $S$, $\hat{U}_{\alpha}(S)$, and the 
quantum-controlled-NOT (QCNOT) on ($S,\mu$), $\hat{U}(S,\mu)$, respectively,
\begin{eqnarray}
&\hat{U}_{2\mu-1} = \hat{U}_{\alpha_{\mu}}(S) = 
\hat{1}(S)\;\cos{(\alpha_{\mu} /2)} - 
\hat{\lambda}_{x}(S)\;i\sin{(\alpha_{\mu} /2)}& \label{us}\\
&\hat{U}_{2\mu} = \hat{U}(S,\mu) = 
\hat P_{00}(S) \hat \lambda_x(\mu) + \hat P_{11}(S) 
\hat{1}(\mu) = \hat{U}^+(S,\mu)&\,. \label{ub}
\end{eqnarray}
However, the basic results of this paper apply also to different 
transformations $\hat{U}(S,\mu)$, e.g. with $\hat{\lambda}_x(\mu)$ replaced by 
$i \hat{\lambda}_y(\mu)$. In any case, the sequence of eq.~(\ref{psi_m}) may 
be interpreted to emerge from a Turing-head 
rotation along the closed Turing-tape, thus iterating between local and 
QCNOT-operations. 
Any such QTM is specified by its tape-size $M$, the external 
{control-parameters 
$\alpha_{\mu}$,} $\mu=1,2, \cdots M$, and the initial {state $|\psi_{0}>$}. 
Without loss of generality we will restrict ourselves 
to $\alpha_{1}=\alpha_{2}= \cdots \alpha_{M}=\alpha$. 
The state $|\psi_{0}>$ will be taken to be 
a product of\, Turing-head and tape wave-functions. This 
initial ``no-correlation'' assumption is typical also for system-bath models 
\cite{MAH95}. In fact, the Turing-tape may be considered 
as a special (finite) bath-model for system $S$. 

We restrict ourselves to the Bloch-vector $\vec{\lambda}$ 
of the Turing-head $S$ (our ``system of interest'')
\begin{equation}
\label{bloch}
\lambda_{i}^{m} =\, <\psi_{m}|\hat{\lambda}_{i}(S) 
\otimes \hat{1}(1) \otimes 
\cdots \otimes \hat{1}(M)|\psi_{m}>\,.
\end{equation}
The Bloch-vectors of the Turing-tape could be calculated 
along the same lines, but the Turing-head 
plays a specific role by construction. Due to the entanglement with the 
Turing-tape, the Turing-head will, in general, appear to be in a ``mixed 
state'', $|\vec{\lambda}^{m}|^{2} < 1$. 

The tape spin-states 
\begin{equation}
|\pm(\mu)>\, = \frac{1}{\sqrt{2}} (|0(\mu)> \pm\, |1(\mu)>)\,,\;\;\;\;\mu=1,2, 
\cdots, M 
\end{equation}
are eigenstates of $\hat{\lambda}_{x}(\mu)$ with eigenvalues $\pm 1$ 
respectively. If spin $\mu$ 
is in one of these states, the QCNOT-operation $\hat{U}(S,\mu)$ cannot create 
any entanglement, irrespective of the head state $|\varphi(S)>$:
\begin{eqnarray}
\hat{U}(S,\mu)\;|\varphi(S)> \otimes\, |+(\mu)> &=& |\varphi(S)> \otimes\, 
|+(\mu)> \\
\hat{U}(S,\mu)\;|\varphi(S)> \otimes\, |-(\mu)> &=& \hat{\lambda}_{z}(S) 
|\varphi(S)> \otimes\, |-(\mu)>\,. 
\end{eqnarray}
For the $2^{M}$ orthonormalized initial tape-states 
\begin{equation}
|{\mathcal{P}}_{0}^{j}>\, \in \{|{\mathcal{P}}_{0}^{\pm\pm \cdots \pm}>\, 
= |\pm(1)> \otimes\, |\pm(2)> \otimes \cdots \otimes\, |\pm(M)>\} \nonumber 
\end{equation}
and with 
$|\varphi_{0}(S)>=\cos(\varphi_{0}/2) |0(S)> -\, i \sin(\varphi_{0}/2) 
|1(S)>$, the network-state $|\psi_{m}>$ remains a product-state, 
\begin{equation}
\label{product}
|\psi_{m}>\,= |\varphi_{m}(S)>\otimes\, |{\mathcal{P}}_{0}^{j}>\;, 
\end{equation}
and the Turing-head described by 
\begin{equation}
\label{bloch_prim}
\lambda_{i}^{m}({\mathcal{P}}_{0}^{j}) =\, 
<{\mathcal{P}}_{0}^{j}|<\varphi_{m}(S)|\, \hat{\lambda}_{i}(S) \otimes 
\hat{1}(1) \otimes 
\cdots \otimes \hat{1}(M)\, |\varphi_{m}(S)>|{\mathcal{P}}_{0}^{j}>
\end{equation}
performs a pure state trajectory on the Bloch-circle
\begin{equation}
\bigl(\lambda_{y}^{m}({\mathcal{P}}_{0}^{j})\bigr)^{2} + 
\bigl(\lambda_{z}^{m}({\mathcal{P}}_{0}^{j})\bigr)^{2} = 1\,.
\end{equation}
We show examples for 
$\varphi_{0} = \pi/6$ and $M=1,2$ (Fig \ref{fig_primitive}). 
The step number $m$ is marked to specify the apparent ``jumping''. The 
explicit machine rules for the Turing-head are given 
in {Table \ref{prit}} ($M=1$). 
The orbits for tape $M$ are contained in those for $k M$ ($k = 2,3, \cdots$). 

For given Turing-tape size $M$ the initial state 
$|{\mathcal{P}}_{0}^{j}>$ 
gives rise to a periodic orbit whose period does not depend on $\alpha$, if
\begin{equation}
|{\mathcal{P}}_{0}^{j}>\, = 
|+>^{n_{0}}|->|+>^{n_{1}}|->|+>^{n_{2}} \cdots 
|->|+>^{n_{q-1}}|->|+>^{n_{q}}\,, \nonumber
\end{equation}
with $n_{i}=0,1,2, \cdots $\,, contains\, $q$ $|->$-states, 
where $\sum_{i=0}^{q} n_{i} + q = M$; 
$q =$ odd for $M$ odd or even, while\, $q$ can be even only 
for $M$ even, satisfying 
\begin{equation}
\sum_{i=0}^{q/2}\,n_{2i}\, =\, \frac{M-q}{2}\,. 
\end{equation}
Otherwise $|\varphi_{m}(S)>|{\mathcal{P}}_{0}^{j}>$ generates an 
aperiodic orbit (i.e. an effective rotation controlled by 
$\alpha$). The aperiodic 
(``quasi-periodic'') primitives also become strictly periodic, if $\alpha$ is 
a rational multiple of $\pi$. 

Any initial state with $S$ in the pure state $|\varphi_{0}(S)>$ can thus be 
written as 
\begin{equation}
\label{q_parallel1}
|\psi_{0}>\, = \sum_{j=1}^{2^{M}}\,a_{j}\,|\varphi_{0}(S)>
|{\mathcal{P}}_{0}^{j}>\,, 
\end{equation}
i.e. $|\psi_{0}>$ can be specified by the coefficients $\{a_{j}\}$. 
With eq.~(\ref{q_parallel1}) and using the orthogonality of the 
$|{\mathcal{P}}_{0}^{j}>$, the resulting motion of the Turing-head depends 
only on the modulus of $a_{j}$ and is given by 
(cf. eqs.~(\ref{bloch}), (\ref{product})) 
\begin{equation}
\label{q_parallel2}
\lambda_{k}^{m} (\psi_{0}) = \sum_{j=1}^{2^{M}}\,|a_{j}|^{2}\,
\lambda_{k}^{m} ({\mathcal{P}}_{0}^{j})\,. 
\end{equation}
This decomposition can be seen as an intuitive example for quantum 
parallelism: The individual Turing-head performs exponentially many primitive 
trajectories ``in parallel''. 
We may restrict the sum in (\ref{q_parallel1}), (\ref{q_parallel2}) 
to the periodic (aperiodic) primitives only. Equal 
weight superpositions of the $4$ periodic ($4$ aperiodic) 
orbits lead to the isolated point (quasi-$1$-dimensional) patterns as shown 
in Fig \ref{3mem} ($M=3$, $\varphi_{0}=0$). The special equal-weight 
superposition with $a_{j} = (1/2^{M})^{1/2}$ corresponds to the initial state 
$|\psi_{0}>\, = |\varphi_{0}(S)> \otimes\, |00 \cdots 0>$, 
which is a complete product-state. There are other non-product 
states though, leading to the same equal-weight result for the Turing-head, 
i.e. to the same pattern. 

For $|\psi_{0}>\, = |00 \cdots 0>$, the typical initial state also 
for quantum computation \cite{DIV95}, and for large $M$ the construction of 
the Turing-head motion based on the 
decomposition approach ($2^{M}$ primitives with equal weight) becomes 
impractical. 
Surprisingly, the Bloch-vector of 
$S$ can easily be found for any $M$ and any step-number $m=n+2M(p-1)$ from 
\begin{eqnarray}
\lambda_{x}^{m} &=& 0 \nonumber \\ 
\lambda_{y}^{m} &=& Y_{m,M}(\alpha) \label{recursion_s}\\
\lambda_{z}^{m} &=& Z_{m,M}(\alpha) \nonumber
\end{eqnarray} 
using the recursion relations (Table \ref{recursion}). Alternatively, the 
Bloch-vector 
$\vec{\lambda}^{m}$ can be calculated directly from the initial state 
\cite{MK98}. The resulting geometrical patterns for $M=1,2,3,10$ are shown 
in {Fig \ref{spindown}}\, including all steps up to\, $m=3000$. 
These patterns, reminiscent of 
Poincar\'{e} sections in classical phase spaces (for open quantum systems 
compare \cite{SPI94}), decompose into various 
sub-manifolds (which reflect higher-order invariants). In the process of their 
built-up the Bloch-vector $\vec{\lambda}^{m}$ jumps between these 
sub-manifolds, just as between the discrete points of the corresponding 
superposition of all the periodic orbits (a one-to-one correspondence, 
compare {Fig \ref{3mem}}); the latter thus play an important role 
reminiscent of Gutzwiller's {\it periodic-orbit-theory} \cite{GUT90}. 
For $M=1,2$ the sub-manifolds are circles 
with radius $r$, center $\vec{c}_{j}$, defining the invariants $I$
\begin{equation}
\label{inv}
I(\vec{\lambda}) = 
\prod_{j=1}^{2^{M+1}} \big(|\vec{\lambda}-\vec{c}_{j}| - r\big) = 0
\end{equation}  
(for $\varphi_{0} = 0$ two of the circles coincide). 
We note in passing that the initial state 
$|10 \cdots 0>$ generates a Turing-head trajectory with 
$\vec{\lambda}^{m}$ of {Fig \ref{spindown}} 
replaced by $-\vec{\lambda}^{m}$ (The individual tape spin may be in any 
state $|0>$, $|1>$). The unitary evolution of a 
mixed state can thus be constructed as weighted combinations of 
these trajectories, at each step $m$. They lead to ``shrunk'' patterns. 

The unitary transformations $\hat{U}_{\alpha}(S)$ and $\hat{U}(S,\mu)$ do not 
commute for $\alpha \ne 0$: Even without introducing any time-parameters, 
the sequence of transformation thus defines a specific order. This ordering 
can be made explicit by associating a {time $\Delta t$} with each step $m$. 
The Fourier transform of this discrete dynamics (underlying the 
built-up of the Turing-head pattern) will thus give complementary information, 
accessible to spectroscopy. This would amount to testing 
the ``non-classicality'' 
of the respective trajectory rather than testing the non-classicality of 
states. Absolute time-scales become relevant as we compare $\Delta t$ with 
the decoherence {time $\tau_{c}$}. Even short {times $\tau_{c}$} 
might be overcome by running the Turing machine fast enough , i.e. by choosing 
$2M \Delta t \ll \tau_{c}$. Note that the Turing-head dynamics is robust 
with respect to phase changes of the Turing-tape states. 


In conclusion, we have shown that the QTM architecture allows for 
a discrete dynamical evolution which, when viewed from the reduced subspace 
of the Turing-head, appears as some highly ordered geometric pattern. 
For specific initial states (``input''), these patterns (``output'') 
can be easily calculated for any tape size. 
They constitute a sensitive local test for the functioning of the total 
network in its exponentially large Hilbert-space. 
The ``output'' becomes available for any large enough observation period 
and does not suffer from the notorious ``halting problem'' \cite{MYE97}. 
These findings, we believe, are the first concrete results pertaining to 
QTM's, a field which up to now has not shown much potential 
for future applications.


We would like to thank C. Granzow, A. Otte and R. Wawer for stimulating 
discussions.


\begin{figure} 
\caption{\label{fig_primitive} 
The primitives ${\mathcal{P}}_{0}^{+}$ (aperiodic) and ${\mathcal{P}}_{0}^{-}$ 
(periodic) for\, $M=1$, and\, ${\mathcal{P}}_{0}^{++}$ (aperiodic), 
${\mathcal{P}}_{0}^{--}$, ${\mathcal{P}}_{0}^{+-}$, 
${\mathcal{P}}_{0}^{-+}$ (periodic) 
for $M=2$; {$\alpha=\pi/\sqrt{3}$},\, {$\varphi_{0}=\pi/6$}.}
\end{figure}

\begin{figure} 
\caption{\label{3mem} 
Equal-weight superpositions ($a_{j}=1/2$) of $4$ periodic ($4$ aperiodic) 
orbits for {$|\psi_{0}>\;=|0000>$},\, {$M=3$},\, and 
total step number $m=3000$. The equal-weight superposition ($1/\sqrt{2}$) of 
these two, in turn, generates the pattern for initial {state $|0000>$} 
(see Fig \ref{spindown} for $M=3$); {$\alpha=\pi/\sqrt{3}$}.}
\end{figure}

\begin{figure}
\caption{\label{spindown} 
Turing-head-patterns for {$|\psi_{0}>\;=|00 \cdots 0>$},\, {$M=1,2,3,10$}, and 
total step number $m=3000$; {$\alpha=\pi/\sqrt{3}$}.} 
\end{figure}

\begin{table}
\caption{State-evolution of Turing-head for $M=1$ and initial states 
$|{\mathcal{P}}_{0}^{j}>$: $\lambda_{y}^{m}({\mathcal{P}}_{0}^{j}) = 
Y_{m}^{(j)}$,\, $\lambda_{z}^{m}({\mathcal{P}}_{0}^{j}) = Z_{m}^{(j)}$,\, 
$j=1$ (aperiodic),\, $2$ (periodic). \label{prit}}
\begin{center}
\begin{tabular}{l|l}
$Y_{0}^{(1)}=\sin{\varphi_{0}}$&$Y_{0}^{(2)}=Y_{0}^{(1)}$\\
$Z_{0}^{(1)}=-\cos{\varphi_{0}}$&$Z_{0}^{(2)}=Z_{0}^{(1)}$\\  
$Y_{1}^{(1)}=Y_{2}^{(1)}=\sin{(\varphi_{0}+\alpha)}$&
$Y_{1}^{(2)}=-Y_{2}^{(2)}=Y_{1}^{(1)}$\\
$Z_{1}^{(1)}=Z_{2}^{(1)}=-\cos{(\varphi_{0}+\alpha)}$&
$Z_{1}^{(2)}=Z_{2}^{(2)}=Z_{1}^{(1)}$\\ 
$Y_{3}^{(1)}=Y_{4}^{(1)}=\sin{(\varphi_{0}+2\alpha)}$&
$Y_{3}^{(2)}=-Y_{4}^{(2)}=-Y_{0}^{(2)}$\\
$Z_{3}^{(1)}=Z_{4}^{(1)}=-\cos{(\varphi_{0}+2\alpha)}$ etc&
$Z_{3}^{(2)}=Z_{4}^{(2)}=Z_{0}^{(2)}$ etc\\
\end{tabular}
\end{center}
\end{table}

\begin{table}
\caption{Recursion relations for the reduced state evolution of $S$ in 
the case of $|\psi_{0}>$ $=|00 \cdots 0>$ and 
$\alpha_1=\alpha_2= \cdots =\alpha_{M}$. Let $Y_{m}=Y_{m,M}$, 
$Z_{m}=Z_{m,M}$, $Z_{m,0}:=-1$, and $m':=m -4p + 2$, where $p$ is the 
cycle number for step $m$; $m=n+2M(p-1)$, $n=1,2, \cdots, 2M$. $Y_{0}=0$, 
$Y_{1}=\sin{\alpha}$, $Z_{0}=-1$, $Z_{1}=-\cos{\alpha}$. \label{recursion}} 
\begin{center}
\begin{tabular}{c|l}
$Y_{m}=-Y_{1}Z_{m-1}-Z_{1}Y_{m-1}$ & $n=$ odd\\
\hline
$Y_{m,M}=Y_{m-1,M}+Y_{1}Z_{m',M-2}$ & $n=$ even $\ne 2M$\\
\hline
$Y_{m,M}=Y_{m-1,M}-Y_{1}(-Z_{1})^{M-1}$ & $n=2M$, $p=$ odd\\
\hline
$Y_{m,M}=Y_{m-1,M}$ & $n=2M$, $p=$ even\\
\hline
$Z_{m} = -Z_{1} Z_{m-1}+ Y_{1} Y_{m-1}$ & $n=$ odd\\
\hline
$Z_{m} = -Z_{1} Z_{m-2}+ Y_{1} Y_{m-2}$ & $n=$ even
\end{tabular}
\end{center}
\end{table}

\begin{figure}
\vspace*{21cm}
\hspace*{-3.5cm}
\includegraphics{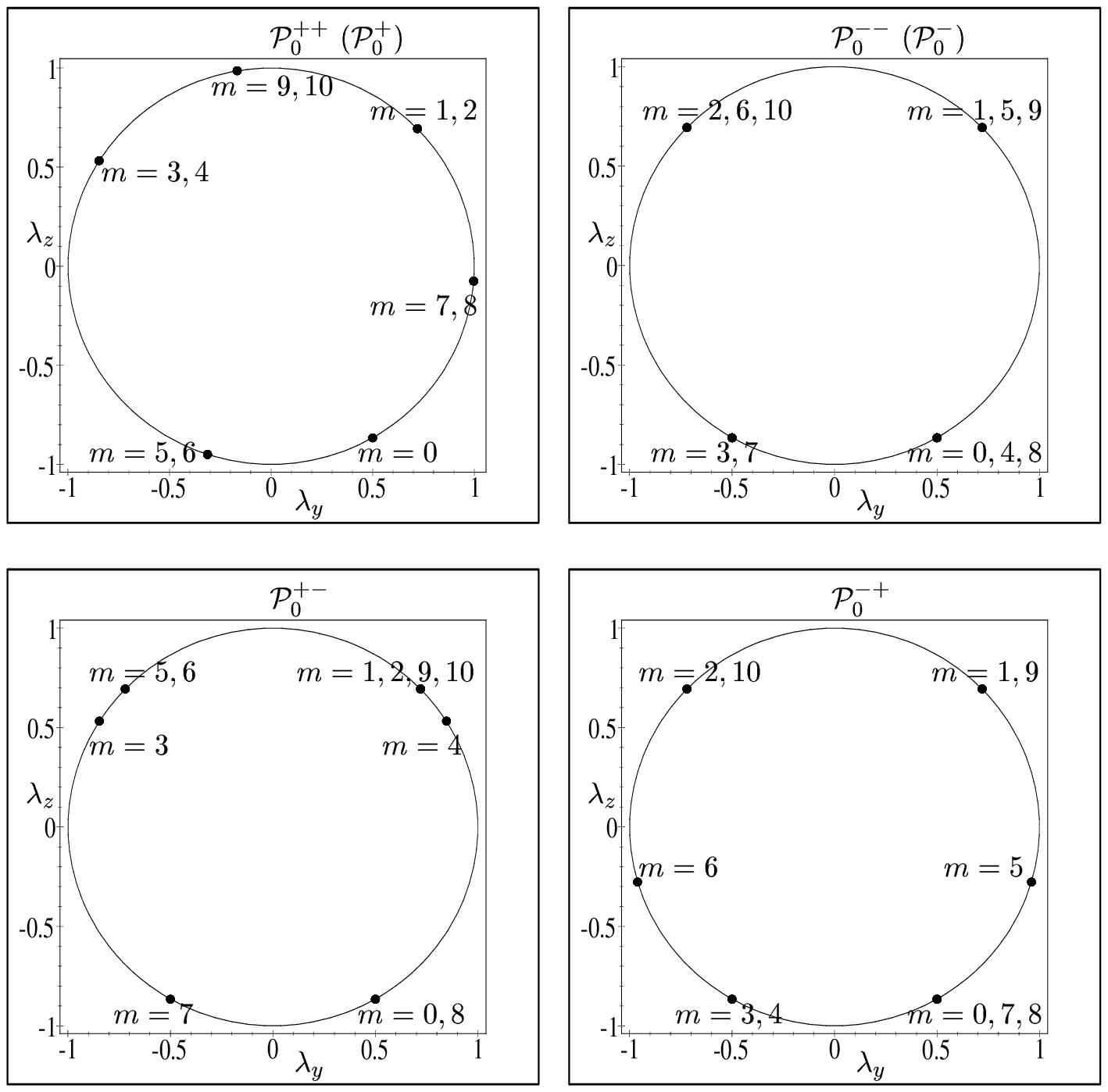}
\end{figure}
\newpage
\begin{figure}
\vspace*{21cm}
\hspace*{-0.25cm}
\includegraphics{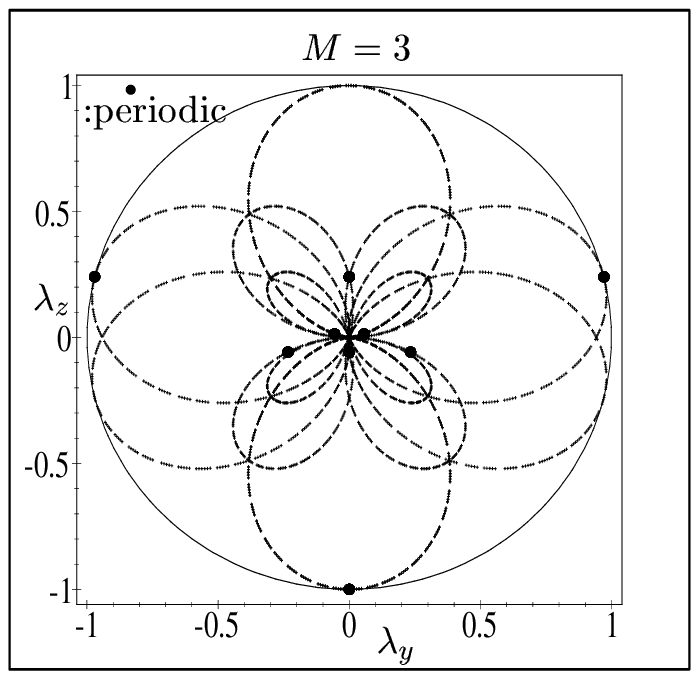}
\vspace*{-10.5cm}
\end{figure}

\begin{figure}
\vspace*{16cm}
\hspace*{-3.5cm}
\includegraphics{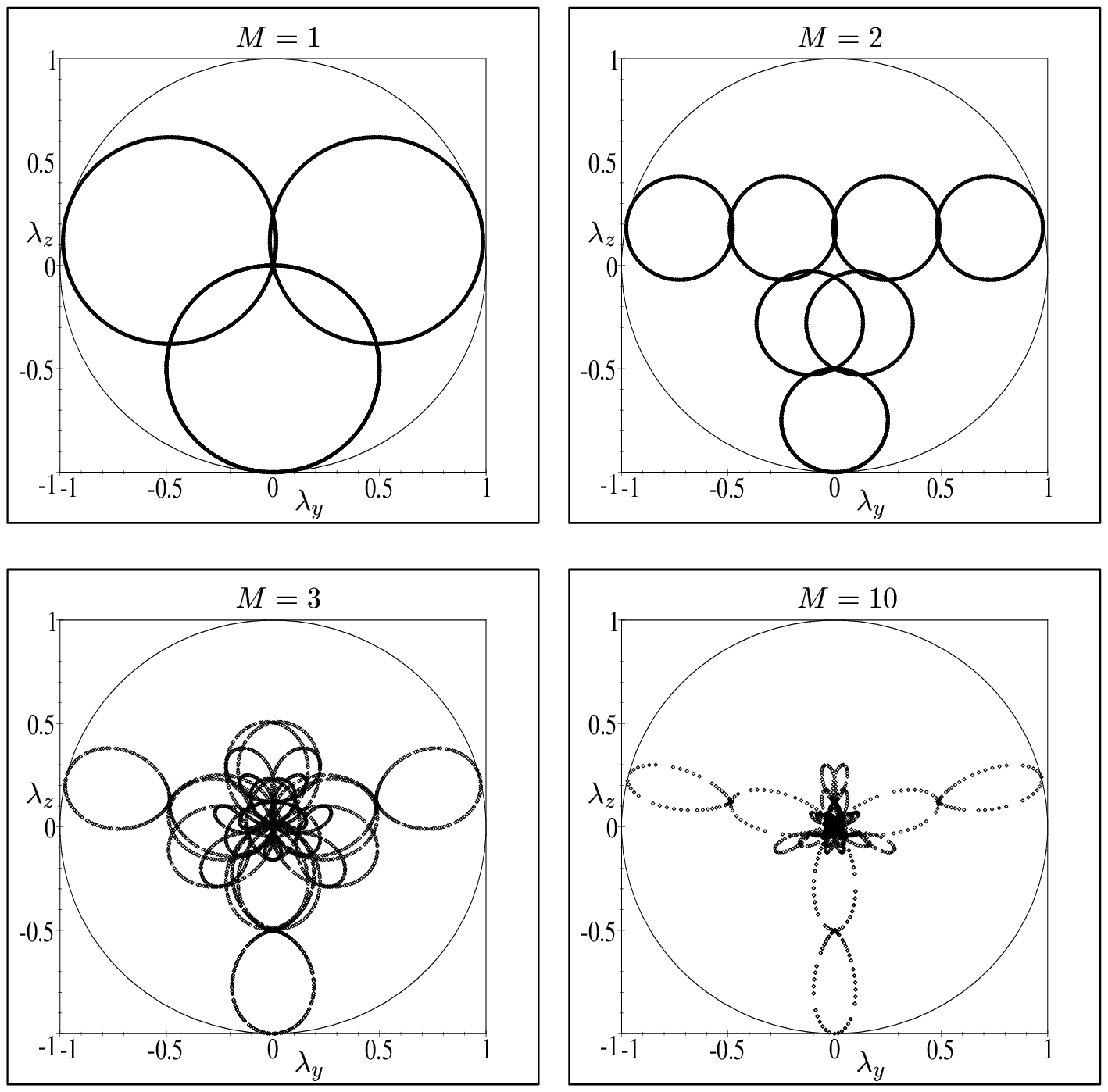}
\vspace*{-7.3cm}
\end{figure}

\begin{thebibliography}{10}
\bibitem{LIC83}
A.~J. Lichtenberg and M.~A. Lieberman, {\em Regular and Stochastic Motion}
  (Springer Verlag, New York, 1983).
\bibitem{HIL94}
R.~C. Hilborn, {\em Chaos and Nonlinear Dynamics} (Oxford University Press, New
  York, 1994).
\bibitem{WOL85}
S. Wolfram, Phys. Rev. Lett. {\bf 54},  735  (1985).
\bibitem{MAH95}
G. Mahler and V.~A. Weberruss, {\em Quantum Networks: Dynamics of Open
  Nanostructures}, 2nd ed. (Springer Verlag, Berlin, New York, 1998).
\bibitem{FEY82}
R.~P. Feynman, Int. J. theor. Phys. {\bf 21},  467  (1982).
\bibitem{DIV95}
D.~P. DiVincenzo, Science {\bf 270},  255  (1995).
\bibitem{EKE96}
A. Ekert and R. Jozsa, Rev. Mod. Phys. {\bf 68},  733  (1996).
\bibitem{LLO96}
S. Lloyd, Science {\bf 273},  1073  (1996).
\bibitem{DEU85}
D. Deutsch, Proc. R. Soc. Lond. A {\bf 400},  97  (1985).
\bibitem{DEU89}
D. Deutsch, Proc. R. Soc. Lond. A {\bf 425},  73  (1989).
\bibitem{BEN82}
P. Benioff, Phys. Rev. Lett. {\bf 48},  1581  (1982).
\bibitem{BEN96}
P. Benioff, Phys. Rev. A {\bf 54},  1106  (1996).
\bibitem{BEN98}
P. Benioff, Fortschr. Physik {\bf 46},  423  (1998).
\bibitem{MK98}
G. Mahler and I. Kim,  in {\em Lecture Notes in Computer Science} (Springer
  Verlag, New York, 1998), in press, also available as online-preprint
  quant-ph/9803008.
\bibitem{SPI94}
T.~P. Spiller and J.~F. Ralph, Phys. Lett. A {\bf 194},  235  (1994).
\bibitem{GUT90}
M.~C. Gutzwiller, {\em Chaos in Classical and Quantum Mechanics} (Springer
  Verlag, New York, 1990).
\bibitem{MYE97}
J.~M. Myers, Phys. Rev. Lett. {\bf 78},  1823  (1997).
\end{thebibliography}
\end{document}